\documentclass[fleqn,usenatbib]{mnras}
\usepackage{xspace,microtype,xcolor}

\usepackage[T1]{fontenc}
\usepackage{ae,aecompl}
\usepackage{hyperref}

\usepackage{graphicx}	
\usepackage{amsmath}	
\usepackage{amssymb}	

\newcommand{\src}{V~0332$+$53\xspace}

\title[Luminosity dependence of the CRSF in \src]{Luminosity dependence of the cyclotron line and evidence for the accretion regime transition in \src}

\author[V. Doroshenko et al.]{Victor Doroshenko$^{1}$, Sergey S. Tsygankov$^{2,5}$, Alexander A. Mushtukov$^{3,4,5}$,
\newauthor Alexander A. Lutovinov$^{5,6}$, Andrea Santangelo$^1$, Valery F. Suleimanov$^1$, Juri Poutanen$^{2,7}$
\\
\\
\normalsize{$^{1}$IAAT, University of Tuebingen, Sand 1, Tuebingen, 72076, Germany}\\
\normalsize{$^2$ Tuorla Observatory, Department of Physics and Astronomy, University of Turku,
 V\"ais\"al\"antie 20, FI-21500 Piikki\"o, Finland} \\
\normalsize{$^3$ Anton Pannekoek Institute, University of Amsterdam, Science Park 904, 1098 XH Amsterdam, The Netherlands} \\
\normalsize{$^4$ Pulkovo Observatory of the Russian Academy of Sciences, Saint Petersburg 196140, Russia}\\
\normalsize{$^5$ Space Research Institute of the Russian Academy of Sciences, Profsoyuznaya Str. 84/32, Moscow 117997, Russia} \\
\normalsize{$^6$ Moscow Institute of Physics and Technology, Moscow region, Dolgoprudnyi, Russia}\\
\normalsize{$^7$ Nordita, KTH Royal Institute of Technology and Stockholm University, Roslagstullsbacken 23, SE-10691 Stockholm, Sweden}\\
}

\date{Accepted XXX. Received YYY; in original form ZZZ}

\pubyear{2016}

\begin{document}
\label{firstpage}
\pagerange{\pageref{firstpage}--\pageref{lastpage}}

\maketitle

\begin{abstract}
We report on the analysis of {\it NuSTAR} observations of the Be-transient
X-ray pulsar \src during the giant outburst in 2015 and another minor outburst
in 2016. We confirm the cyclotron-line energy -- luminosity correlation
previously reported in the source and the line energy
decrease during the giant outburst. Based on
2016 observations we find that a year later the line energy has
increased again essentially reaching the pre-outburst values. We discuss this behaviour and
conclude that it is likely caused by a change of the emission region geometry
rather than previously suggested accretion-induced decay of the neutron stars magnetic field. 
At lower luminosities we find for the first time a hint of departure from the anti-correlation of line energy with
flux, which we interpret as a transition from super- to sub- critical accretion
associated with disappearance of the accretion column. Finally, we confirm
and briefly discuss the orbital modulation observed in the outburst light curve
of the source.
\end{abstract}


\section{Introduction} 
In binary systems accretion of matter supplied by non-degenerate companion onto
a strongly magnetised ($B\sim10^{12}$\,G) rotating neutron star results into
pulsed X-ray emission from the vicinity of neutron stars (NSs) magnetic poles. The plasma is
channeled to the polar caps by the magnetic field of the neutron star which has
also profound effect on the observed X-ray spectra. In particular, the motion
of the electrons in strong magnetic field is quantised, which gives rise to the
so-called cyclotron resonance scattering features (CRSFs, see
\cite{Mushtukov15x} for a recent review). A single (fundamental) or multiple
harmonics \citep{Santangelo99} absorption-like features can be observed in
X-ray band depending on the conditions in the line forming region. In particular, the energy of
the fundamental is related to the magnetic field strength as $E_{\rm cycl}\sim 12\mbox{keV}\,B/10^{12}$\,G.

The structure of the emission region in the vicinity of the NS and thus the
origin of the CRSF are, however, uncertain. At low accretion rates most of the
observed emission likely comes directly from the accretion mounds on the polar
caps of the NS where the gravitational energy of the flow is released. However,
the observed luminosities of bright pulsars by far exceed the local Eddington
limit for kilometre-sized polar caps. At high accretion rates, the plasma must
be, therefore, stopped above the NS surface by the radiative pressure and the
observed emission has to emerge from the extended ``accretion column''
\citep{Basko,Becker12,2015MNRAS.454.2539M}. Conditions for the transition
between the two regimes are determined by the largely unknown geometry of the
column, and by the angular- and energy-dependent plasma opacities which makes
it extremely hard to make robust theoretical predictions on the transition (critical)
luminosity \citep{Mushtukov15,2015MNRAS.454.2539M}.

On the other hand, analysis of the luminosity dependence of the observed
properties of X-ray pulsars might help to constrain the critical luminosity
observationally \citep{Tsygankov06,Staubert07,Klochkov12}. Indeed, in low
luminous sources the CRSF energy typically increases with the flux, whereas at
higher accretion rates an anti-correlation is observed. As discussed by
\cite{Staubert07}, \cite{Becker12}, \cite{Mushtukov15}, and
\cite{Mushtukov15pcor}, this behaviour could point on the two accretion regimes
corresponding to sub- and super-critical accretion. Observing the transition
between the two regimes in a single source would strongly support this
interpretation. In this paper we report on the analysis of the CRSF luminosity
dependence in the \emph{Be-}transient X-ray pulsar \src during the giant
outburst in 2015 and another minor outburst in 2016, and discuss the complex evolution of the line
energy throughout the giant outburst and between the two outbursts which, we argue,
provides the first evidence for such transition.

\section{Observations and data analysis} 
\begin{figure}
   \centering
      \includegraphics[width=\columnwidth]{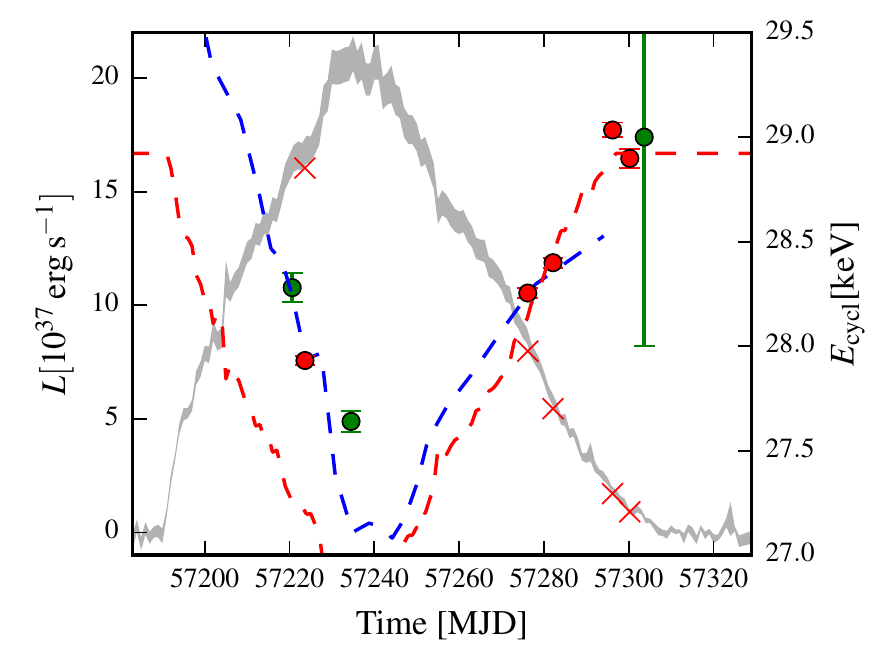}
   \caption{{\it Swift}/BAT light curve in 15--50\,keV band (gray) of \src
   during the 2015 giant outburst scaled to match the source luminosity
   measured using {\it NuSTAR} pointed observations (red crosses). The red
   and green circles with error bars indicate the CRSF fundamental energy as measured by
   {\it NuSTAR} and {\it INTEGRAL}/SPI respectively. The red dashed line
   shows the model prediction for the fundamental energy based on the CRSF
   energy versus luminosity correlation measured by {\it NuSTAR} in the
   declining part of the outburst. The blue dashed line shows the
   fundamental energy reported by \protect\cite{Cusumano16} shifted by 1\,keV.
   The shift is likely due to the difference in absolute energy scale of the {\it Swift}/BAT with respect to {\it NuSTAR} and SPI.}
   \label{fig:overview}
\end{figure}

\begin{figure}
   \centering
      \includegraphics[width=\columnwidth]{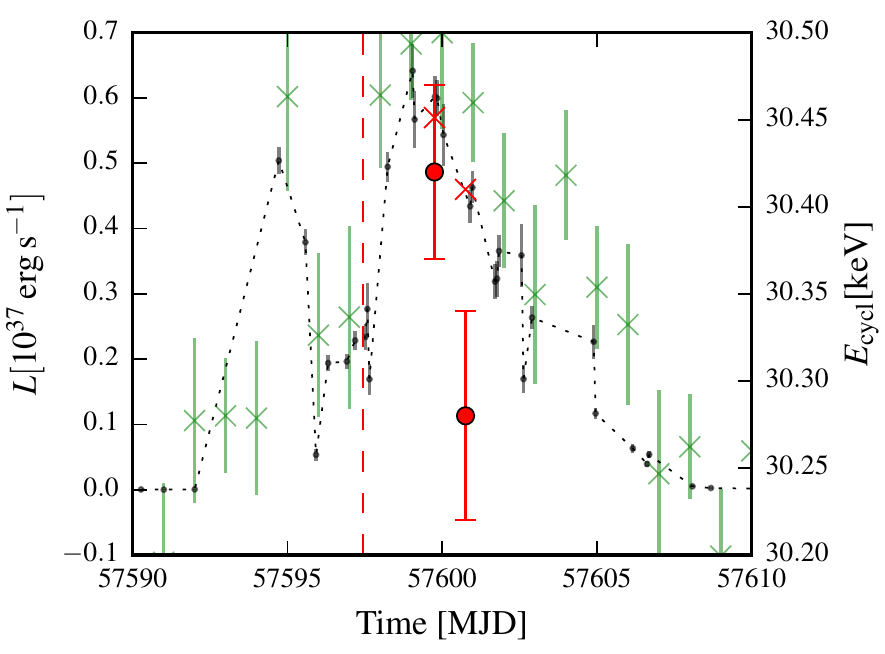}
   \caption{{\it Swift}/XRT (0.3--10\,keV, gray points with error bars) and
   {\it Swift}/BAT (15--50\,keV, green crosses with error bars) light-curves of \src
   during the Aug 2016 periastron passage scaled to match the source luminosity
   measured using {\it NuSTAR} pointed observations (red crosses). The red
   error bars indicate the CRSF fundamental energy as measured by
   {\it NuSTAR}.}
   \label{fig:overview_low}
\end{figure}

We focus on the analysis of five
dedicated {\it NuSTAR} observations during the 2015 giant outburst aimed to
detect the transition from super- to sub-critical accretion regime. We verify
the {\it NuSTAR} results using the {\it INTEGRAL}/SPI observations of the
source during the outburst ({\it INTEGRAL} revolutions 1565, 1570, and 1596).
Additionally, we report on two follow-up {\it NuSTAR} observations obtained in
Jul 2016 during another periastron passage aimed to extend the observational
coverage at low luminosities. We note, that the source flux in the last two
observations approaches that for the transition of the source to propeller
\citep{Tsygankov16}, so we probe here the lowest fluxes when the source is
still accreting. The observation log is presented in Table~\ref{tab:results}
and Figs.~\ref{fig:overview} and \ref{fig:overview_low}. Finally, we used also
{\it Swift}/XRT data contemporary to the {\it NuSTAR} observations to extend
the low-energy band and monitor the activity of the source near the periastron
passage in July 2016.

The {\it NuSTAR} data reduction was carried out using the HEASOFT~6.19 package
with current calibration files (CALDB version 20160824) and standard
data reduction procedures as described in the instruments documentation. Source
spectra were extracted from a region of 120$^{\prime\prime}$ radius around the
\src. The background spectra were extracted from a circular region of
80$^{\prime\prime}$ radius as far away from the source as possible for each
observation. The spectra for the two {\it NuSTAR} units were extracted and
fitted simultaneously between 5 and 79\,keV for the outburst observations
\citep{Fuerst13}, and in 3-79\,keV range for the last two observations where no
Swift data was available. To extract the {\it Swift}/XRT spectra we used the
Swift data products service provided by the UK Swift Science Data
Centre\footnote{\url{http://www.swift.ac.uk/user_objects/}} as described in
\cite{Evans09}. All {\it NuSTAR/Swift} spectra were grouped to at least
25 counts per bin and fit using the
{\sc xspec} version 12.9 with Gehrels weighting \citep{Gehrels86}.

{\it INTEGRAL} observed \src\ four times during the spacecraft revolutions 1565,
1570, 1586 and 1596. Problems with the energy calibration of IBIS and JEM-X
telescopes did not permit us to reconstruct the source spectrum properly.
However, we were able to do it for three observations where the SPI
spectrometer was operating (detectors annealing was performed during revolution
1586). The {\it INTEGRAL}/SPI data were screened and reduced in accordance with
the procedures described by \cite{2011MNRAS.411.1727C,2014Natur.512..406C}. 

The broadband spectrum of the source has been previously described
\citep{Tsygankov06,Lutovinov15} using a power law with cutoff at high energies
(\texttt{CUTOFFPL} model in {\sc xspec}) modified by interstellar absorption
and one to three broad absorption features accounting for the CRSF at
$\sim26$\,keV and its harmonics at $\sim50$\,keV and $\sim72$\,keV
\citep{Tsygankov06}. To account for the CRSF and the first harmonic we use the
multiplicative gaussian $G(E)=1-D_{\rm cycl} e^{-\ln{2}((E-E_{\rm
cycl})/\sigma_{\rm cycl})^2}$ rather than exponential gaussian ({\texttt GABS}
in Xspec), or the pseudo-lorentzian model (\texttt{CYCLABS} in Xspec) used by
\cite{Tsygankov06}. Indeed, the later model was designed to mimic the high
energy cutoff \citep{Mihara90} which is already included in the continuum
model. As a consequence, the measured line centroid $E_0$ becomes coupled to
the cutoff energy and shifted by $\sigma^2/E_0$ with respect to true centroid
\citep{Nakajima10}, which complicates interpretation of the results. On the
other hand, exponential gaussian profile usually used to describe the CRSFs
yields a slightly worse fit with systematic $\sim$1-2\% residuals around the
line, especially for \texttt{CUTOFFPL} continuum. This behaviour has been
reported by \cite{Pottschmidt05} and \cite{Nakajima10} for 2005 outburst and
was interpreted as evidence for a complex CRSF profile. We find, however, that
the magnitude of the residuals depends on the continuum model used (for
instance, they essentially disappear for \texttt{HIGHECUT} model). Furthermore,
restricting the energy range to 20--80\,keV as well as using multiplicative
gaussian or lorentzian line profile results in no significant residuals for any
continuum model. We conclude, therefore, that given the existing uncertainties
in modelling of the broadband continuum of X-ray pulsars, there is no strong
evidence for a more complex line profile in \emph{NuSTAR} data. This conclusion
is consistent with Swift/BAT results \citep{Cusumano16} where gaussian line
provided adequate description of the data.

\begin{figure}
   \centering
      \includegraphics[width=\columnwidth]{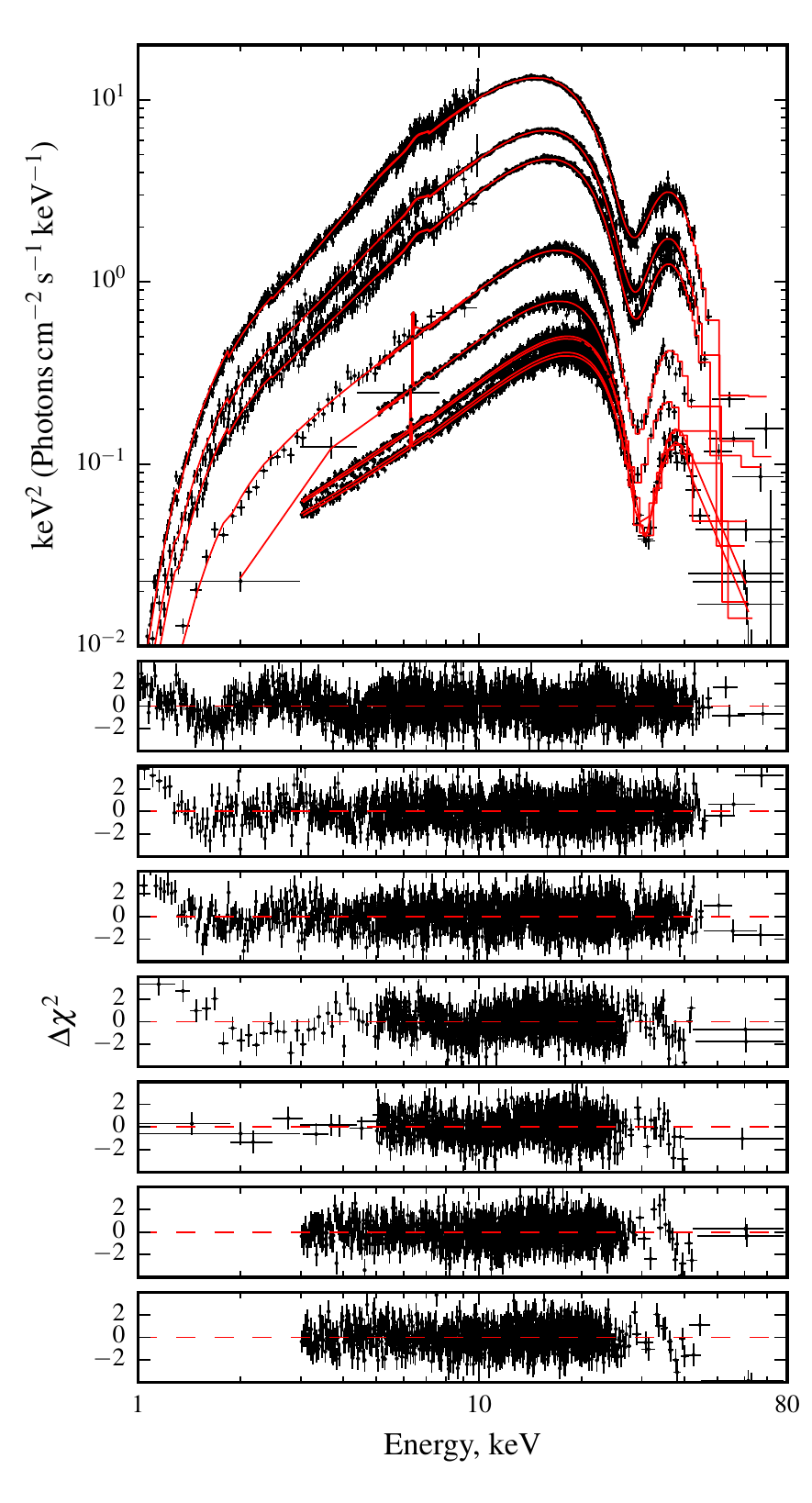}
   \caption{Unfolded spectra as observed by {\it NuSTAR} (5--80\,keV) and {\it Swift}/XRT (1--10\,keV),
    and the corresponding best-fit residuals for the brightest to the dimmest observations (top to bottom, top panel).}
   \label{fig:spectrum}
\end{figure}

We verified that measured CRSF centroid does not depend on the continuum or
line model used and is well constrained for all {\it NuSTAR} observations. In
particular, we measured consistent CRSF energies (within the uncertainties)
using the broadband fits of {\it NuSTAR}+{\it Swift}/XRT data and
\texttt{HIGHECUT} or a comptotnisation model {\texttt CompTT} by
\cite{Titarchuk}, as well as for {\it NuSTAR} data in 20--80\,keV range and the
\texttt{CUTOFFPL} model for either lorentzian and gaussian line profiles. In
all cases inclusion of the additional soft blackbody component with temperature
of $\sim0.4$\,keV improves the fit for the XRT data, although, taking into
account large systematic uncertainties in window-timing mode it is unclear
whether this component is real. For all models we also accounted for
interstellar absorption. It was sufficient to assume the absorption column
fixed to the average value of $2\times10^{22}$\,cm$^{-2}$ for all observations.
We note that the absorption column is similar to one derived from XRT
observations in later phases of the outburst \citep{Tsygankov16}, so there is
no evidence for enhanced absorption during the bright phase of the outburst.
Neither component significantly affects the derived CRSF parameters. The
\texttt{CompTT} continuum model provides, however, the most stable and
consistent fit for all observations, therefore, we use this model combined with
the gaussian profile for the CRSF for the rest of analysis. On the other hand,
{\it INTEGRAL} data does not allow to reliably constrain the continuum, so we
fix respective parameters to the values derived from the {\it NuSTAR} data at
closest luminosity and only fit for the CRSF parameters. Again, we have
verified that the derived line parameters are not significantly affected by
choice of the continuum model also in this case. We also found that at low
fluxes the width of the first harmonics becomes poorly constrained due to the
correlation with continuum temperature. We assumed, therefore, that the
relative width of the harmonic (i.e $\sigma_{\rm cycl,1}/E_{\rm
cycl,1}=\sigma_{\rm cycl}/E_{\rm cycl}$) is the same at all luminosities. In
spectra with highest statistical quality the second harmonic becomes visible in
the residuals, although not really significant. This is not surprising as the
{\it NuSTAR} is only sensitive up to 78\,keV, so we did not include the second
harmonic in the model. The results of the fits are presented in
Table.~\ref{tab:results}. Unfolded spectra for the best-fit model and
respective residuals are shown in Fig.~\ref{fig:spectrum}. All uncertainties
are quoted at 1$\sigma$ confidence level and include no systematic error unless
stated otherwise.

\begin{figure}
   \centering
     \includegraphics[width=\columnwidth]{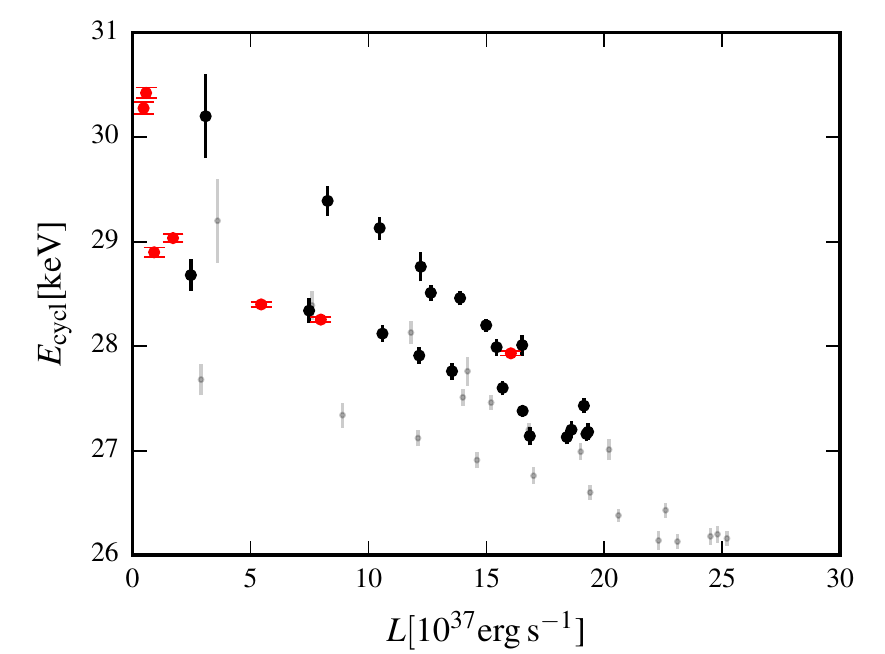}
   \caption{Dependence of the fundamental centroid energy on luminosity as
   observed by {\it NuSTAR} in 2015-2016 (red points). The same correlation as
   reported by \protect\cite{Cusumano16} based on {\it Swift}/BAT data is also plotted
   for reference (gray points). Note that the assumed spectral model and energy range
   used to calculate the luminosity was different in two cases (see text). The
   apparent $\sim1$\,keV shift between the two instruments is likely related to
   difference in their absolute energy calibration. Once this shift and
   correction to the luminosity are taken into the account, the agreement between
   {\it Swift} and other instruments becomes acceptable (black points).}
   \label{fig:corr}
\end{figure}

It is interesting to note that there is an apparent shift by $\sim 1$\,keV
between {\it NuSTAR} and {\it INTEGRAL} data (which are consistent with each
other) and the {\it Swift}/BAT measurements reported by \cite{Cusumano16} as
illustrated in Fig.~\ref{fig:corr}. This mismatch is probably related to the
difference in absolute energy calibration between the instruments as {\it
INTEGRAL} measurements are also consistent for the current and previous
outbursts \citep{Ferrigno16}. Besides this shift, the difference in absolute
flux calibration of the instruments and difference in energy ranges used to
calculate luminosities needs to be taken into the account for direct comparison
of the results. In particular, we re-calculated the {\it Swift}/BAT
luminosities reported by \cite{Cusumano16} using the {\it Swift}/BAT
15$-$50\,keV light-curve and contemporary {\it NuSTAR} fluxes which turn out to
be well correlated. Based on this correlation we estimate
$L_x=91(2)C$[10$^{37}$erg\,s$^{-1}$] conversion factor (here $C$ is observed
BAT count-rate). Once said corrections are taken into the account, the CRSF
energies measured by all three observatories become compatible within
uncertainties as illustrated in Fig.~\ref{fig:corr}.


\begin{figure}[ht]
    \centering
        \includegraphics[width=\columnwidth]{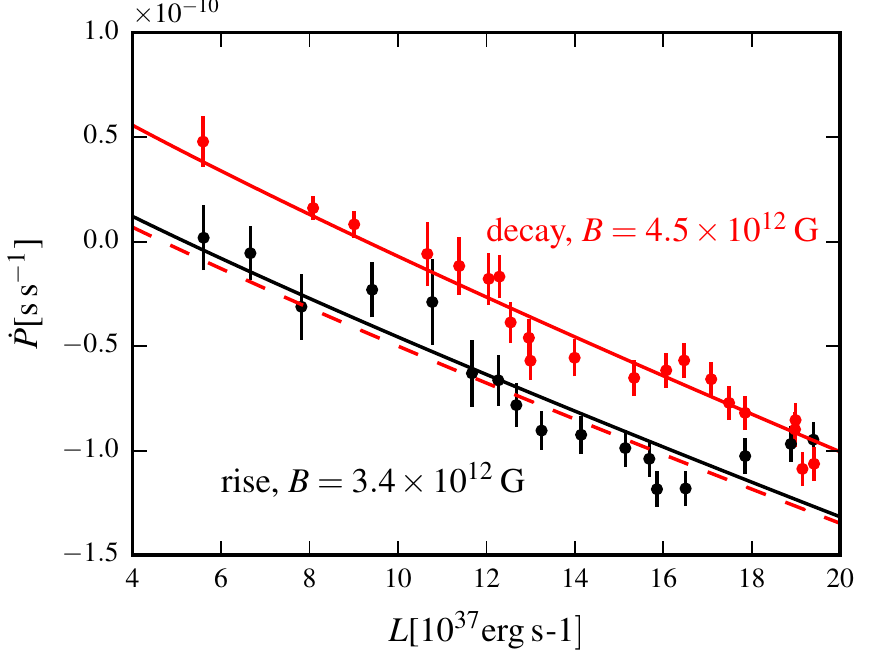}
    \caption{Correlation of the spin-up rate with flux as reported in \protect\cite{Doroshenko16}.
    A best-fit prediction for torque model by \protect\cite{Lipunov82a} with $B=3.4\times10^{12}$\,G and $B\sim4.5\times10^{12}$\,G
    for rising (black) and declining (red) parts of the outburst respectively is also shown for reference. 
    A $\sim5$\% magnetic field decay suggested by \protect\cite{Cusumano16} would imply slightly higher spin-up rate for the declining part of the outburst 
    (with respect to the rising part) as indicated with red dashed line.}
    \label{fig:spinev}
\end{figure}
\begin{figure*}
   \centering
      \includegraphics[width=0.33\textwidth]{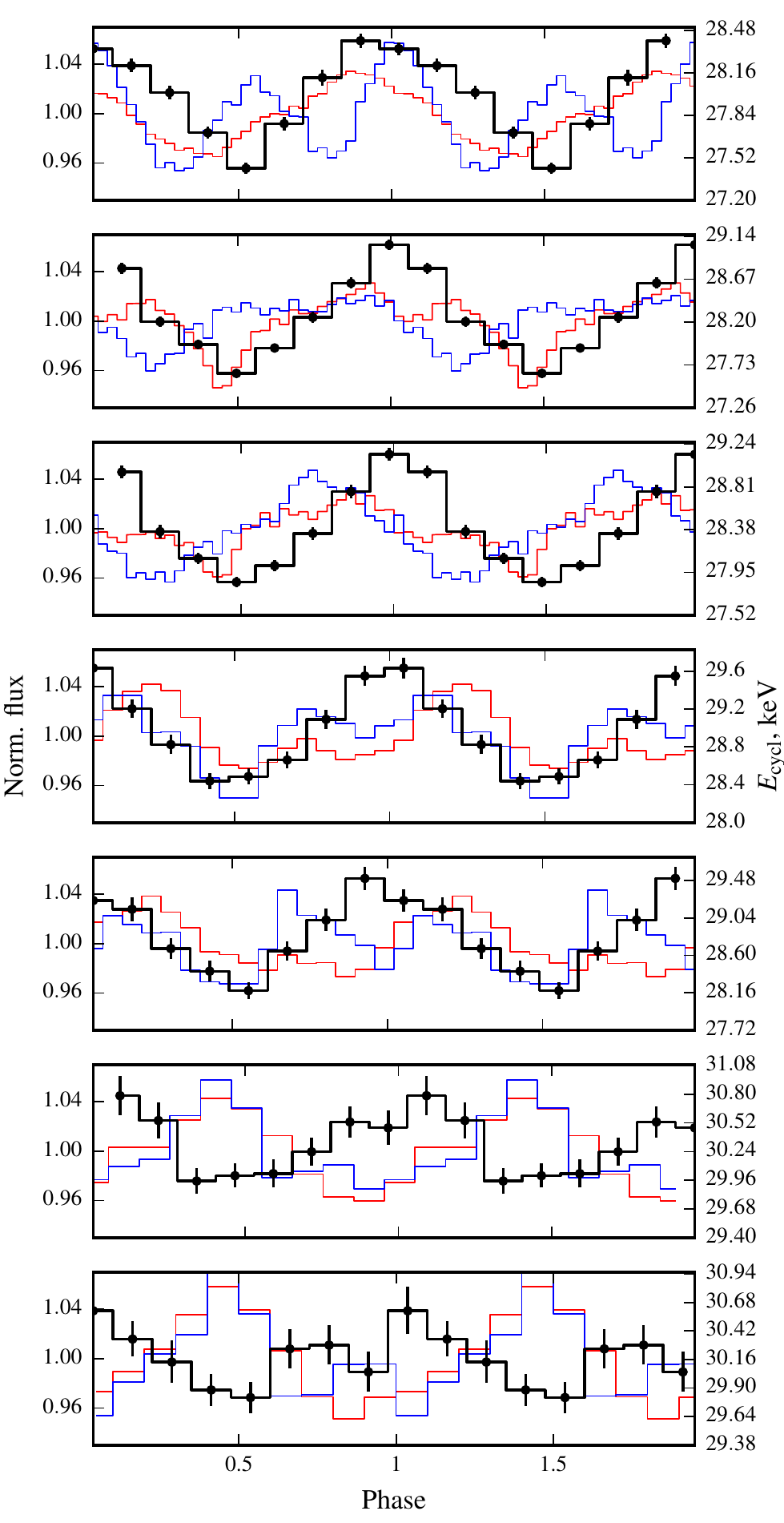}
      \includegraphics[width=0.33\textwidth]{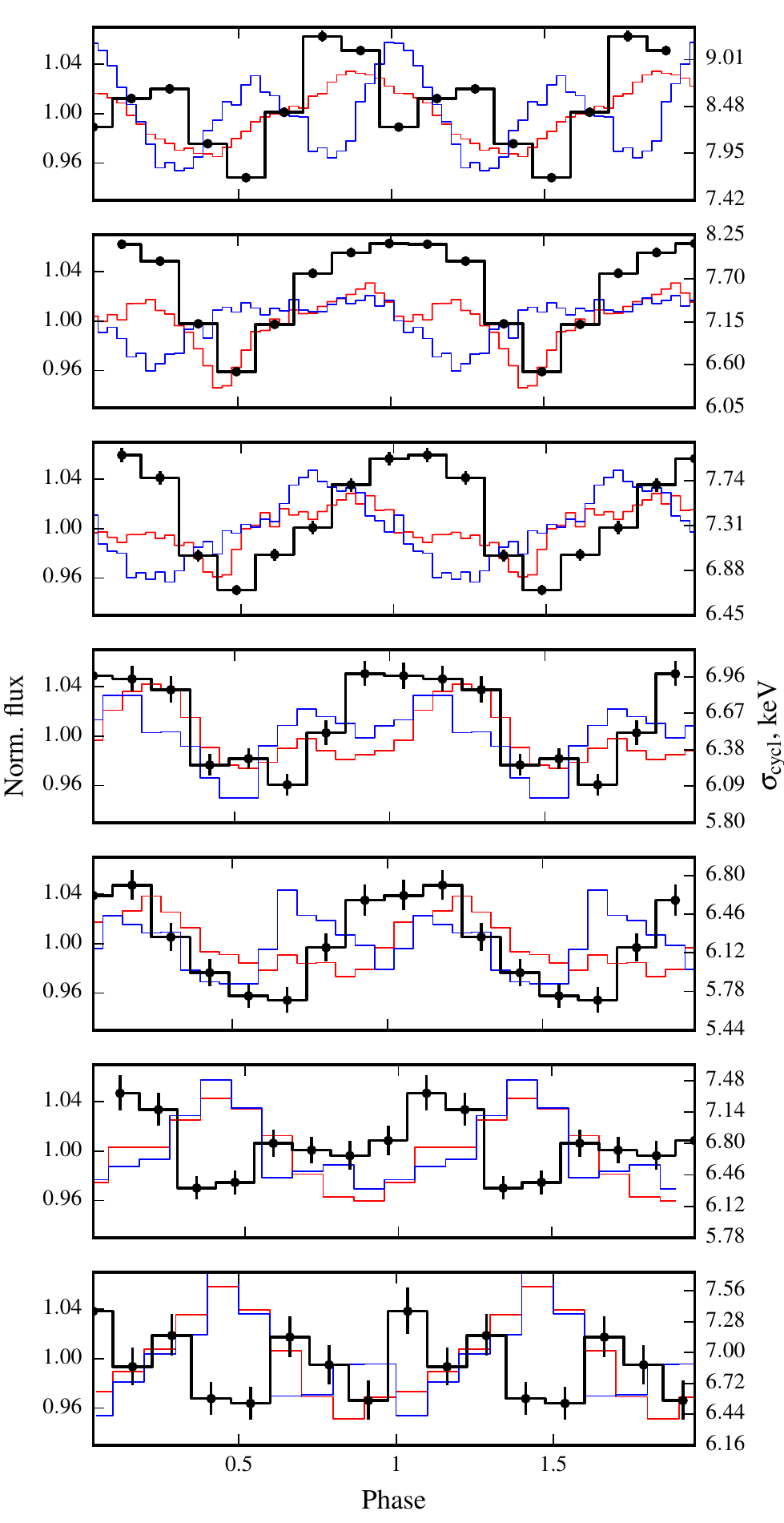}
      \includegraphics[width=0.33\textwidth]{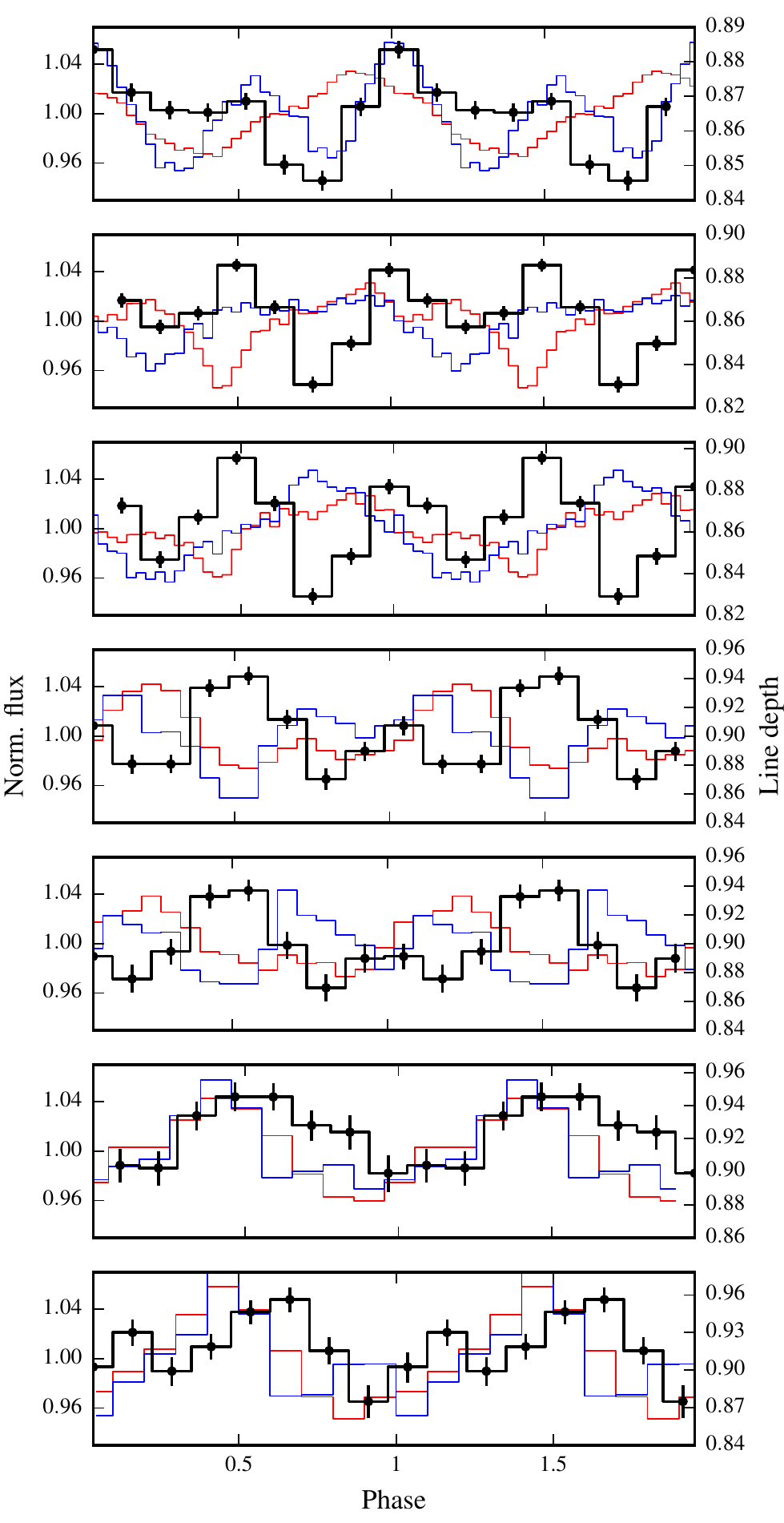}
   \caption{Pulse phase dependence of the CRSF energy as observed by {\it
   NuSTAR} (black histogram) at different luminosities
   ($L/10^{37}$\,erg\,s$^{-1}$=16.04, 7.98, 5.45, 1.71, 0.91, 0.57, 0.46 top to bottom).
   The pulse profiles in 3--15 and 15--80\,keV range (blue and red
   steps) are also shown. The pulse phase of individual observations was
   aligned so that zero phase corresponds to maximal CRSF energy.}
   \label{fig:phaseres}
\end{figure*}
\begin{figure}
   \centering
      \includegraphics[width=\columnwidth]{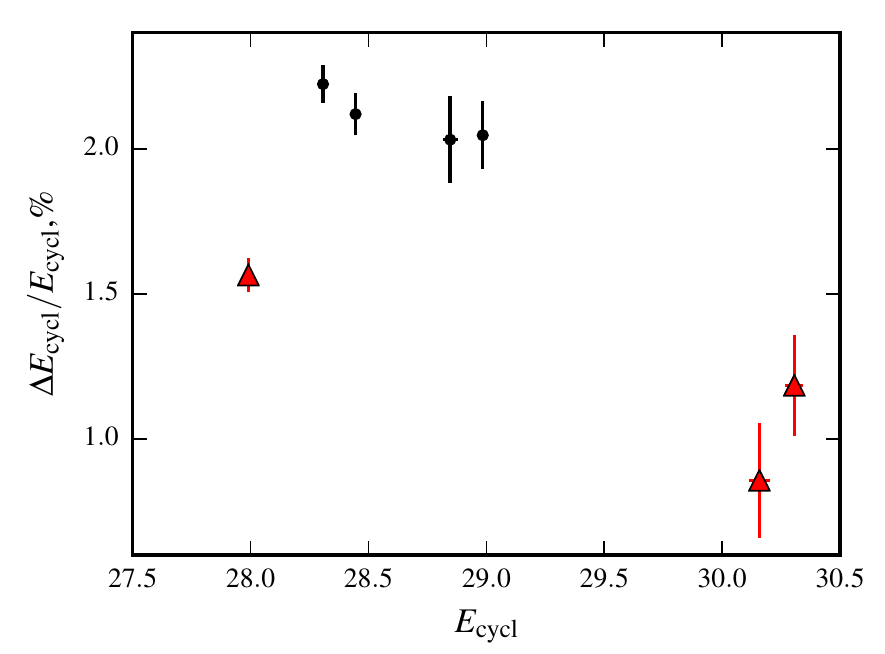}
   \caption{Amplitude of variation of the fundamental centroid energy with
   pulse phase for the declining part of the main outburst (black points) and
   for the rest of the data (red triangles).}
   \label{fig:ecycampl}
\end{figure}
\begin{figure}
      \centering
            \includegraphics[width=\columnwidth]{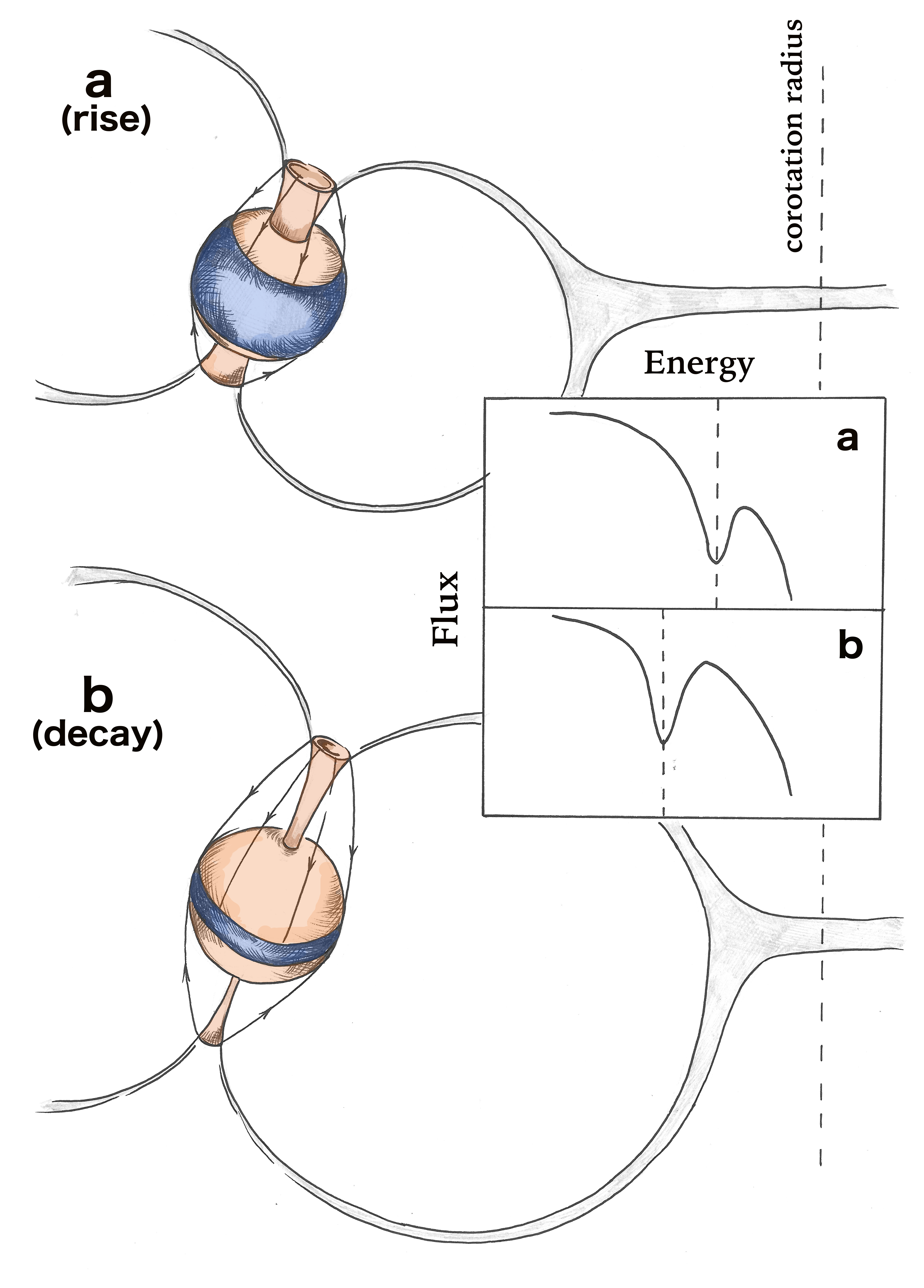}
      \caption{Change of the accretion disc structure and of the magnetospheric
      radius (with respect to the co-rotation radius indicated by vertical line) 
      change both the net torque exerted onto the NS, and the accretion column height.
      During the declining phase the matter falls closer to the polar areas which increases the height
      of the accretion column so that illumination pattern shifts to the equatorial areas implying lower
      observed CRSF energy.}
      \label{fig:art}
\end{figure}
\begin{figure}
   \centering
     \includegraphics[width=\columnwidth]{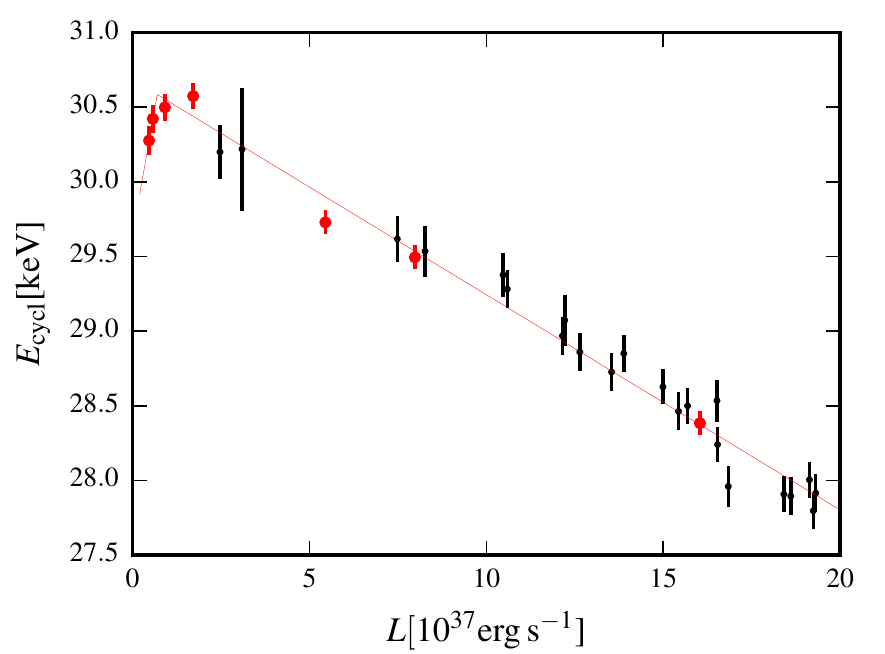}
   \caption{Correlation of the CRSF centroid energy on luminosity with the linear drift
   of $\dot{E}_{\rm cycl}=-0.015$\,keV/d taken into the account (symbols are the same as in Fig.~\protect\ref{fig:corr}). Note the transition from anti-correlation
   to correlation below $10^{37}$erg\,s$^{-1}$.}
   \label{fig:corr_adj}
\end{figure}
\begin{figure} \centering 
\includegraphics[width=\columnwidth]{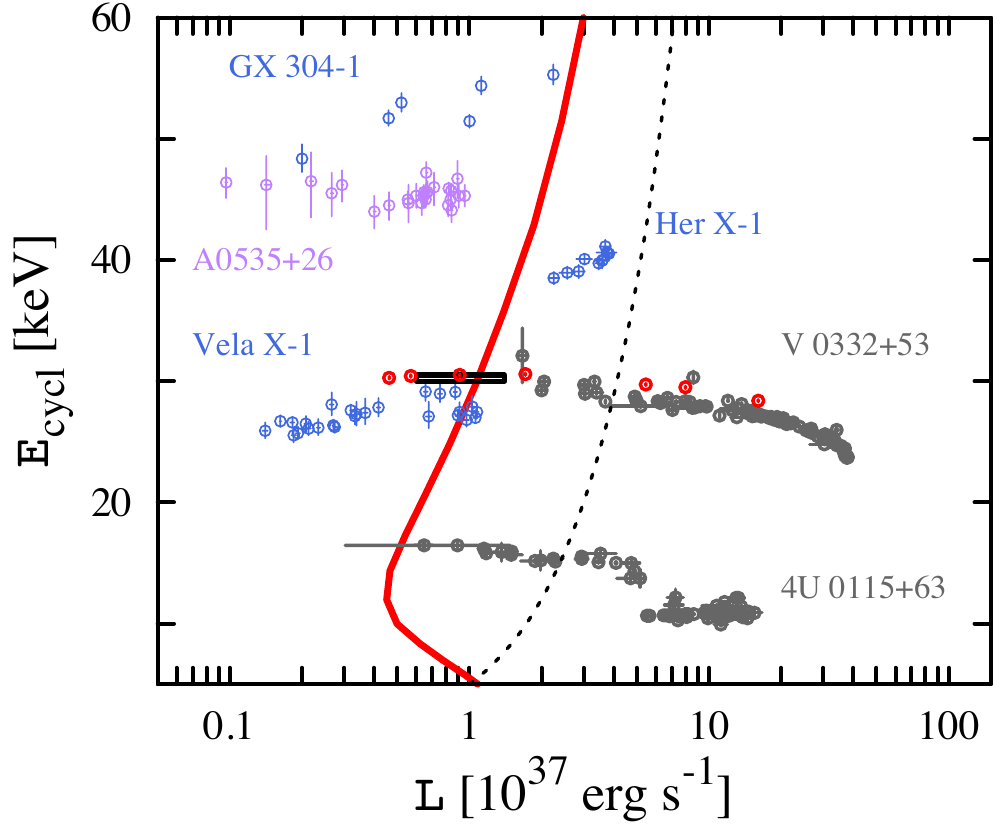}
\caption{CRSF energy as function of luminosity for sources where either a
correlation or anti-correlation was reported in the literature. The red circles correspond to {\it NuSTAR}
data reported in this work. The black box corresponds to the estimated value of
the transitional luminosity. The red solid curve corresponds to the critical
luminosity value \protect\citep{Mushtukov15} for the case of $\Lambda=0.5$, where $\Lambda$ is the radio of
magnetospheric radius to Alfv\'en radius, and radiation dominated by $X$-mode
polarisation.} \label{fig:lxcrit}
\end{figure}

We also carried out the pulse-phase resolved analysis of all {\it NuSTAR}
observations. The complex spin frequency evolution \citep{Doroshenko16} and
long intervals between the individual observations prevented us from obtaining
a single phase-coherent timing solution. Therefore, we phased all observations
using the reference epochs obtained via fitting the CRSF energy pulse profile
with a cosine function. The corresponding pulse profiles in 3--15 and
15--80\,keV energy bands are also plotted for reference in
Fig.~\ref{fig:phaseres}. For the phase-resolved analysis we additionally fixed
the \texttt{CompTT} temperatures and iron line parameters to average values.

\begin{table*}
	\centering
	\caption{Observation log and best-fit results
   for the phase averaged spectrum using the \texttt{CompTT} model.
   For INTEGRAL/SPI some of the parameters (shown in italic) were fixed to values derived from {\it NuSTAR} data
   at closest luminosity level. Luminosity is calculated based on the observed flux in 3--80\,keV band not accounting for interstellar absorption or the CRSFs, and assuming distance of 7\,kpc.}
	\label{tab:results}
	\begin{tabular}{lcccccccccc} 
        		\hline
                                 Obs. ID. & Date & Exposure & $E_{\rm cycl}$ & $\sigma_{\rm cycl}$ & $D_{\rm cycl}$ &    $T_0$ &  $kT_e$ &   $\tau$ &    $L_{37}$, 3--80\,keV & $\chi^2$/dof\\
                                     &  MJD &    ks &      keV &        keV &         &     keV &    keV &       & $10^{37}$\,erg\,s$^{-1}$ &       \\
        \hline
                               80102002002 & 57223    &   10.5 &   27.93(2) &       8.47(8) &     0.863(2) &   1.40(1) &  5.85(4) &   16.4(1) &           16.04 &  0.97/2864\\
                               80102002004 & 57276    &   14.9 &   28.25(2) &       7.59(8) &     0.861(2) &   1.19(1) &  5.73(5) &   18.3(1) &           7.98 &  0.9/2628\\
                               80102002006 & 57282    &   17.3 &   28.40(2) &       7.43(8) &     0.862(2) &   1.03(2) &  5.76(5) &   19.7(1) &           5.45 &  0.84/2600\\
                               80102002008 & 57296    &   18.2 &   29.03(4) &       6.73(8) &     0.899(3) &   0.78(3) &  6.08(5) &   18.5(1) &           1.71 &  0.88/1724\\
                               80102002010 & 57300    &   20.1 &   28.90(5) &       6.30(1) &     0.895(4) &   0.73(5) &  6.09(7) &   17.7(1) &           0.91 &  0.83/1375\\
                               90202031002 & 57599.8  &   25.2 &   30.42(5) &       6.80(1) &     0.925(4) &   0.66(5) &  6.80(1) &   16.5(1) &           0.57 &  0.88/1369\\
                               90202031004 & 57600.8  &   25.0 &   30.28(6) &       6.90(1) &     0.921(4) &   0.68(5) &  6.90(1) &   16.4(1) &           0.46 &  0.91/1330\\
                               rev. 1565   & 57220    &   67.8 &   28.28(7) &   \emph{8.47} &     0.845(5) &\emph{1.4} &\emph{5.85} & \emph{16.4}   &         $\sim19.5$    &   0.89/110\\
                               rev. 1570   & 57233    &   141. &   27.64(5) &   \emph{8.47} &     0.814(4) &\emph{1.4} &\emph{5.85} & \emph{16.4}   &         $\sim23.1$    &   1.2/110\\
                               rev. 1596   & 57302    &   120. &   29(2)    &   \emph{6.30}  &     0.4(3)  &\emph{0.73}&\emph{6.1} & \emph{17.7} &          $\sim1.0$    &   0.88/10\\
        
\hline
	\end{tabular}
\end{table*}

\section{Discussion}
\subsection{CRSF centroid energy drift}
The source is known to exhibit an anti-correlation of the CRSF centroid energy
with luminosity which is believed to be caused by change of the accretion
column height and was studied extensively during the 2004-2005 outburst
\citep{Tsygankov06,2010MNRAS.401.1628T, Poutanen13,Lutovinov15}. The
\emph{NuSTAR}, \emph{INTEGRAL} and \emph{Swift} observations reveal similar
behaviour for the current outburst, although with two important differences.
First, line energies measured during the declining part of the outburst seem to
be significantly lower for comparable luminosity levels \citep{Cusumano16}
which was not the case in 2005 \citep{2010MNRAS.401.1628T}. The reported drop
of the centroid energy reaches $\sim1.5$\,keV for {\it Swift}/BAT data (the
blue dashed line in Fig.~\ref{fig:overview}). This is fully consistent with our
{\it NuSTAR} and {\it INTEGRAL} results as shown in Figs.~\ref{fig:overview}
and \ref{fig:corr}. Note that the two {\it NuSTAR} observations in July 2016
reveal, for the first time, that after the $\sim1.5$\,keV drop during the 2015
outburst the centroid energy has again increased by same amount effectively
negating the observed decay during the outburst.

\cite{Cusumano16} argued that the observed decrease of the CRSF energy is due
to the screening of the NSs dipole magnetic field by the accreting matter. The
efficiency of this mechanism depends on the ratio of timescales for field
``burial'' through the advection by accreting matter and re-emerging of the
field driven by magnetic buoyancy, Ohmic diffusion, or other mechanisms
\citep{Choudhuri02}. While both timescales are highly uncertain, it is still
interesting to compare \src with other sources where a variation of the CRSF
energy with time has been reported. The net decay rate of the CRSF during the
outburst is $\dot{E}_{\rm cycl}\sim1.5{\rm keV}/100{\rm\,d}$, and the net
increase rate between the outbursts at least $\dot{E}_{\rm cycl}\sim1.5{\rm
keV}/300{\rm\,d}$. If accretion is indeed responsible for screening of the
magnetic field, the field-increase rate shall remain the same also in outburst,
so the advection shall reduce the field even faster than observed. The
corresponding field decay/re-emerging timescales $\tau=E_{\rm
cycl}/\dot{E}_{\rm cycl}\sim4-16$\,yr turn out to be significantly shorter than
observed for other sources and generally expected from theoretical point of
view. For instance, in Vela~X$-1$ and Her~X$-$1 the CRSF decay rates of
$\sim-9.7\times10^{-4}$\,keV/d \citep{parola16} and
$\sim-7\times10^{-4}$\,keV/d \citep{Klochkov15,Staubert16} imply
$\tau\sim70-155$\,yr.

Another issue with this interpretation is related to spin evolution of \src
during the outburst which is governed by balance of the accelerating and
braking torques exerted onto the neutron star (\citealt{Rappaport77},
\citealt{Ghosh79}, \citealt{Lipunov81}, \citealt{Lipunov82}, \citealt{Wang87}).
While both torques increase with the magnetosphere size, the braking torque is
more sensitive to the field strength and thus shall decrease faster than the
accelerating torque if the intrinsic field of the neutron star indeed drops by
$\sim5$\% as suggested by \cite{Cusumano16}. One can, therefore, expect the net
spin-up rate of the neutron star to increase during the declining phase of the
outburst. However, the opposite is observed as reported by \cite{Doroshenko16}
and illustrated in Fig.~\ref{fig:spinev}. The observed spin-up rate actually
decreases for a given accretion rate during the later phases of the outburst
which implies $\sim30$\% increase of the magnetic field strength (assuming the
torque model by \cite{Lipunov82a} and $B=3.4\times10^{12}$\,G field for the
rising phase of the outburst as deduced from the observed CRSF energy,
however, similar result can be obtained for \cite{Ghosh79} model). We
conclude, therefore, that spin evolution of the source is inconsistent with the
intrinsic field decay suggested by \cite{Cusumano16} and must be related to
other factors. Detailed discussion of this issue is out of scope for this work,
and we can only speculate that the observed change of the spin-up rate might be
associated with viscous evolution of the accretion disc during the outburst.
Indeed, the disc is expected to have higher surface density and thus might push
further into the magnetosphere during the rising phase of the outburst which
would imply lower magnetospheric radius (for the same field strength of the
neutron star) and thus higher spin-up rate. We note that comparable total
amount of accreted mass during 2005 and 2015 outbursts \citep{Cusumano16}
together with significantly longer duration of the later outburst suggests that
the accretion disc did indeed have different structure in two cases.

So what besides the decay of intrinsic magnetic field could cause the observed
CRSF energy decrease? We note that the magnitude of the line energy drop is
comparable with the variation of line energy with pulse phase (see
Fig.~\ref{fig:phaseres}) and luminosity, so it is natural to attribute the
observed CRSF decay to a change in the geometrical configuration of the line
forming region. Such change can be caused by several factors besides the
variation of NSs intrinsic field, for instance by a change of the effective
magnetospheric radius, which, as discussed above, is also suggested by spin
evolution of the pulsar. In context of model by \cite{Poutanen13}, the CRSF is
formed via reflection of beamed radiation from the accretion column off the
unevenly illuminated NS atmosphere, so the observed change in CRSF energy
corresponds to a change of the illumination pattern. The footprint of the
accretion column is expected to be reduced for larger magnetospheric radii as
the plasma follows the field lines which are closer to the magnetic pole in
this case. Decrease of the footprint implies that the accretion column becomes
taller for a given luminosity and thus more effectively illuminates equatorial
regions of the NS resulting in lower observed CRSF energy as illustrated
schematically in Fig.~\ref{fig:art}.

Changes in the emission region geometry must be reflected in pulse profile
shape. Indeed, the pulse profiles observed at comparable luminosities in the
declining phase of the 2015 outburst and in 2016, do appear to be significantly
different as illustrated in Fig.~\ref{fig:phaseres}. On the other hand, in
context of the reflection model almost entire NS surface is illuminated in both
cases, so no drastic changes for phase dependence of the CRSF parameters is
expected which is again qualitatively consistent with the results of phase
resolved analysis. Additional argument supporting the proposed interpretation
comes from the comparison of the relative amplitude of the CRSF energy
variation with pulse phase in different observations. This turns out to be
significantly higher for observations in the declining phase of the outburst
where the illuminated area is larger, and thus larger fraction of the NSs
atmosphere contributes to the line formation as illustrated in
Fig.~\ref{fig:ecycampl}. We note that such change indicates a significant
change in the emission region structure regardless on assumed model for the
CRSF formation.

\subsection{The critical luminosity} 
At high luminosities the CRSF centroid energy in \src is known to be
anti-correlated with flux, which is consistent with \cite{Cusumano16} findings
and our results for the declining phase of the outburst. However, the
anti-correlation seems to break at the lowest flux (see Fig.~\ref{fig:corr}).
Indeed, the centroid energy is well constrained in the last two {\it NuSTAR}
observations in 2015, and is actually slightly lower during the dimmer
observation. Comparison of the observed line energies in two dimmest
observations in 2015 implies thus a positive correlation with luminosity with
$dE/dL=0.16(8)$\,keV/10$^{37}$\,erg\,s$^{-1}$. For the two observations in
2016, one can deduce $dE/dL=1.3(7)$\,keV/10$^{37}$\,erg\,s$^{-1}$, i.e. in four
out of seven \emph{NuSTAR} observations the line energy seems to increase with
flux, i.e. the anti-correlation reported for higher fluxes does not seem to
extend to low fluxes indefinitely. Transition from an anti-correlation to a
correlation is, in fact, expected from theoretical point of view. The
anti-correlation observed at high fluxes is thought to be associated with the
change of height of the accretion column which is supported by radiative
pressure and thus appears only above certain critical luminosity \citep{Basko}.
Below the critical flux the line is expected to be correlated with the flux
either due to the Doppler effect \citep{Mushtukov15pcor} or due to a change of
the atmosphere height above the NS surface driven by ram pressure of the
in-falling material \protect\citep{Staubert07}.

Assuming that such a transition does indeed take place, the transitional
luminosity can be estimated by fitting a broken linear model to \emph{NuSTAR}
and BAT data corrected for the linear drift as shown in
Fig.~\ref{fig:corr_adj}. This yields 
$L_{\rm crit}=0.7_{-0.1}^{+0.7}\times10^{37}$\,erg\,s$^{-1}$ with 
$E_{\rm cycl}=30.58(7)-0.144(4)(L-L_{\rm crit})/10^{37}$ above the transitional luminosity, and 
$E_{\rm cycl}=30.58(7)+1.4(1.2)(L-L_{\rm crit})/10^{37}$ below it. 
Here we
additionally include in quadrature a systematic uncertainty of 0.1\,keV for
BAT\protect\footnote{http://heasarc.nasa.gov/docs/heasarc/caldb/swift/docs/bat/SWIFT-BAT -CALDB-ESCALE-v1.pdf}, and 0.077\,keV for {\it NuSTAR} respectively to
get a statistically acceptable fit (in the later case the systematics
corresponds to the uncertainty of energy scale around the mean CRSF energy
assuming the long-term gain variations reported by \cite{Madsen15}). The
best-fit statistics improves from $\chi^2_{\rm red}=1.37$ for 27 degrees of
freedom for a linear fit to $\chi^2_{\rm red}=1.03$ for 25 degrees of freedom
for the broken linear fit, which implies that the later model is marginally
preferred at $\sim3\sigma$ confidence based on the MLR test
\citep{Protassov02}. Note that while the transitional luminosity value seems to
be in good agreement with theoretical predictions as illustrated in
Fig.~\ref{fig:lxcrit}, both the significance level and the deduced parameters
might be affected by the assumptions that the line energy decreased linearly
during the giant outburst and has fully recovered between the two outbursts.

On the other hand, the slope of the anti-correlation at high fluxes,
which shall be less sensitive to either assumption, is in good agreement with
value reported by \cite{2010MNRAS.401.1628T}, i.e. seems to be robustly
constrained. We can, therefore, compare it with slopes deduced for the low flux
observations as reported above, which implies deviation of $\sim3.5\sigma$ and
$\sim2\sigma$ for 2015 and 2016 observations respectively. Here we include no
systematical uncertainties as the \emph{NuSTAR} gain is expected to remain
stable on short timescales. One can also estimate the correlation slope at low
luminosities by fitting all four low flux observations with a linear model with
a common slope and arbitrary intercepts for 2015 and 2016 low flux
observations. This allows to account for the possibility that the line energy
has not completely recovered between the outburst as well as for possible
long-term energy scale variations. The best-fit
$dE/dL=0.17(8)$\,keV/10$^{37}$\,erg\,s$^{-1}$ and $\sim3.8\sigma$ deviation
from the value obtained for higher fluxes.

We conclude, therefore, that \emph{NuSTAR} data provides a first hint of the
transition from anti-correlation to correlation of the CRSF energy with flux at
low fluxes. As illustrated in Fig.~\ref{fig:lxcrit}, the transitional
luminosity is in agreement with theoretical predictions for onset of an
accretion column, so it is natural to associate the change in correlation slope
with disappearance of the accretion column. On the other hand, comparatively
low statistical significance of the correlation break together with the complex
evolution of line energy throughout the giant outburst and between the
2015 and 2016 outbursts makes it hard to justify whether the transition is
indeed robustly detected, so additional observations are required to confirm
our findings.

\subsection{The ``mini''-outburst in 2016}
Finally, we would like to comment briefly on the {\it Swift}/XRT light curve of
the source in July-August 2016. Traditionally outbursts of Be-transients are
classified in two types, i.e. ``giant '' ones like that observed from \src in
2005, 2015, and ``normal'' ones which typically occur at periastron and are
detectable with all-sky monitors like {\it Swift}/BAT. Both burst types are
evident in long-term \src light curve. However, closer inspection reveals also
some flux enhancements at almost every periastron with level below that typical
for ``normal'' outbursts. This prompted us to request additional {\it NuSTAR}
observations and {\it Swift} XRT monitoring which indeed revealed a minor
outburst with peak luminosity $\sim6\times10^{36}$\,erg\,s$^{-1}$ barely
detectable also by BAT. We note that even lower-level accretion can occur in
\src and other Be-transients also when they are not detected by all-sky
monitors, and this might have important consequences for studies dedicated to
cooling of the neutron stars \citep{Wijnands16}.

The light curve itself is also quite interesting as it reveals a sharp dip in
vicinity of the periastron. Similar behaviour was reported recently by
\cite{Ferrigno16} during the 2015 giant outburst. They interpreted the observed
orbital modulation as flux enhancement following the periastron passage
associated with the accretion of matter captured at periastron and delayed by
propagation through the accretion disc. In our case it is, however, clear, that
the outburst starts before the periastron and the accretion rate drops for
$\sim3$\,d to restore later to the pre-dip level (see
Fig.~\ref{fig:overview_low}). We note that light curve modeling by
\cite{Ferrigno16} is not unambiguous and the observed modulation during the
giant outburst can be also attributed to the dips at periastron rather than
flux enhancement afterwards. We conclude, therefore, that taking into the
account our findings, the interpretation or orbital modulation suggested by
\cite{Ferrigno16} is probably not correct. On the other hand, the observed flux
drop at periastron is at odds with the commonly accepted picture of outbursts
in Be-transients which are believed to be triggered by the enhanced accretion
as the neutron star passes through the disc of the primary close to the
periastron.

Complex outburst development has been reported also for other Be-systems
\citep{Postnov08,Klochkov11}. For 1A~0535+262 \cite{Postnov08} attributed the
initial flare to unstable accretion triggered by some magnetospheric
instability, and the following dip to the depletion of the inner disc regions.
In case of the \src, however, similar behaviour was observed also during the
outburst at high accretion rates, which makes this explanation unlikely. The
dip in \src is also clearly not related to enhanced absorption as the ratio of
XRT/BAT fluxes remains constant. We hesitate to provide a better explanation,
and the discussion of physical origin of dips origin is out of scope of this
paper. Still, we wanted to bring this issue up to illustrate that exploration
of the properties of the Be-transients at low luminosities enabled for the
first time by {\it NuSTAR} and {\it Swift}/XRT is very interesting indeed and
shall be continued.

\section{Conclusions} 
Based on the analysis of {\it NuSTAR} observations of \src during the declining
part of the 2015 giant outburst we have confirmed the previously known
anti-correlation of CRSF energy with luminosity. We also confirm the apparent
drop of the CRSF centroid energy during the declining part recently interpreted
by \cite{Cusumano16} as the result of the accretion-induced decay of the
magnetic field of the NS. We find that line energy decrease is consistent with
being time linear throughout the outburst with rate of $\sim0.015$\,keV/d.
Furthermore, follow-up \emph{NuSTAR} observations of another outburst in 2016
revealed that the line energy has increased again approximately to values
observed before the 2015 outburst which implies a recovery rate of
$\sim0.05$\,keV/d. Both timescales imply unprecedentedly fast evolution of the
magnetic field of the neutron star if the change of the observed line energy is
directly related to field strength as suggested by \cite{Cusumano16}. We argue,
however, that evolution of the observed CRSF energy is likely instead
associated with a change of the emission region geometry. The later is defined
by the magnetosphere size which indeed seems to be different in rising and
declining parts of the outburst as sugested by the observed spin evolution of
the neutron star.

Finally, we find that at luminosities below $\sim10^{37}$\,erg\,s$^{-1}$ the
anti-correlation of the CRSF energy with flux reported for higher luminosities
seems to break, which we interpret as the first observational evidence for the
transition from super- to sub-critical accretion. The transitional luminosity
is in agreement with the theoretical predictions and cyclotron line luminosity
dependence observed in other sources as shown in Fig.~\ref{fig:lxcrit}. We
note, however, that taking into the account complex evolution of line energy
throughout the outburst and comparatively low statistical significance of the
break, the transition can not be considered to be robustly detected and
additional observations are required to confirm our findings.

\section*{Acknowledgements}
\footnotesize Authors thank E.M. Churazov who developed the INTEGRAL/SPI data
analysis methods and provided the software. This research has made use of data
provided by HEASARC (NASA/GSFC and Smithsonian Astrophysical Observatory). This
work is based on observations with INTEGRAL, an ESA project with instruments
and science data centre funded by ESA member states (especially the PI
countries: Denmark, France, Germany, Italy, Switzerland, Spain), and with the
participation of Russia and the USA. This work made use of data supplied by the
UK Swift Science Data Centre at the University of Leicester. The authors
acknowledge support from Deutsches Zentrum f{\"u}r Luft- und Raumfahrt (DLR)
through DLR-PT grant 50\,OR\,0702 (VD), Deutsche Forschungsgemeinschaft (DFG)
through WE 1312/48-1 (VS), the Russian Science Foundation through grant
14-12-01287 (AAM, SST, AAL), the Academy of Finland through grant 268740 and
the Foundations' Professor Pool, the Finnish Cultural Foundation (JP).

\bibliography{biblio}
\bibliographystyle{mnras}

\bsp   
\label{lastpage}
\end{document}